\documentclass[aps,prb,twocolumn,showpacs,superscriptaddress]{revtex4}
\usepackage{graphicx}
\begin{document}

%\preprint{}

\title{Bose-Einstein Condensation of Triplons in Ba$_3$Cr$_2$O$_8$}
     \author{A.~A.~Aczel}
     \altaffiliation{author to whom correspondences should be addressed: E-mail:[aczela@mcmaster.ca]}
     \affiliation{Department of Physics and Astronomy, McMaster University, Hamilton, Ontario, Canada, L8S 4M1}
     \author{Y.~Kohama}
     \affiliation{National High Magnetic Field Laboratory, Los Alamos National Laboratory, Los Alamos, New Mexico 87545, USA}     
     \author{M.~Jaime}
     \affiliation{National High Magnetic Field Laboratory, Los Alamos National Laboratory, Los Alamos, New Mexico 87545, USA} 
     \author{K.~Ninios}
     \affiliation{Department of Physics, University of Florida, Gainesville, Florida 32611, USA}
     \author{H.~B.~Chan}
     \affiliation{Department of Physics, University of Florida, Gainesville, Florida 32611, USA}
     \author{L.~Balicas} 
     \affiliation{National High Magnetic Field Laboratory, Tallahassee, Florida 32310, USA}
     \author{H.~A.~Dabkowska}
     \affiliation{Brockhouse Institute for Materials Research, McMaster University, Hamilton, Ontario, Canada, L8S 4M1} 
     \author{G.~M.~Luke}
     \affiliation{Department of Physics and Astronomy, McMaster University, Hamilton, Ontario, Canada, L8S 4M1}
     \affiliation{Brockhouse Institute for Materials Research, McMaster University, Hamilton, Ontario, Canada, L8S 4M1} 
     \affiliation{Canadian Institute of Advanced Research, Toronto, Ontario, Canada, M5G 1Z8}

\date{\today}

\begin{abstract}
By performing heat capacity, magnetocaloric effect, torque magnetometry and force magnetometry measurements up to 33~T, we have 
mapped out the T-H phase diagram of the S~$=$~1/2 spin dimer compound Ba$_3$Cr$_2$O$_8$. We found evidence for field-induced 
magnetic order between H$_{c1}~=~$12.52(2)~T and H$_{c2}~=~$23.65(5)~T, with the maximum transition temperature T$_c~\sim~$2.7~K 
at H~$\sim$~18~T. The lower transition can likely be described by Bose-Einstein condensation of triplons theory, and this is 
consistent with the absence of any magnetization plateaus in our magnetic torque and force measurements. In contrast, the nature 
of the upper phase transition appears to be quite different as our measurements suggest that this transition is actually first order.  
\end{abstract}

\pacs{
73.43.Nq, %Quantum phase transitions 
75.30.Kz, %Magnetic phase boundaries
75.30.Sg, %Magnetocaloric effect 
75.40.Cx %Critical point effects, specific heat, short-range order: static properties
}
%\keywords{}
\maketitle

%para 1

Quantum phase transitions (QPT) can be achieved by varying a non-thermal control parameter, such as pressure or applied 
magnetic field, while at a temperature of absolute zero\cite{book_sachdev, 03_vojta}. These transitions are driven by 
quantum fluctuations resulting from the uncertainty principle, as opposed to the thermal fluctuations that drive classical 
phase transitions. A particular type of QPT is realized in a Heisenberg spin dimer system, which possesses a non-magnetic 
spin-singlet ground state with a gap to the first triplet excited state\cite{08_giamarchi}. The excited triplets (triplons) can 
be considered as bosons with a hard core on-site repulsion\cite{02_rice}. The repulsion condition is necessary in order to 
prevent more than one triplon from lying on a single dimer. 

If one applies a magnetic field H to close the spin gap, a critical field H$_{c1}$ is eventually reached which 
results in the generation of a macroscopic number of triplons. Above H$_{c1}$, the magnetic field can be varied to 
control the triplon density, and so it acts as a chemical potential. The system now consists of a series of interacting 
triplons with a ground state that critically depends on the balance between the kinetic energy and the repulsive
interactions. Note that the kinetic energy of the interacting triplons arises from the xy-component of the Heisenberg 
interdimer interaction, while the nearest neighbour repulsive interaction (different from the on-site repulsion) arises 
from the Ising or z-component. The delicate balance between these two energies has led to interesting and diverse 
properties of QPTs in spin dimer systems.

If the repulsive interactions dominate, it is most crucial to minimize this contribution to the microscopic 
Hamiltonian. The easiest way to do this is to ensure that the triplon density per dimer is a simple rational fraction, as 
this allows the triplons to form a superlattice. These preferred fractional triplon densities result in plateaus in
the magnetization as a function of field, and such behaviour has been observed in SrCu$_2$(BO$_3$)$_2$ (Ref.~[5,6]).

When the kinetic energy terms dominate instead, this contribution will be minimized by allowing the triplons to have 
freedom to hop from dimer to dimer. The ground state then consists of a coherent superposition of singlets and
triplets. No magnetization plateaus are observed in this case, but rather there is a continuous rise in the magnetization 
from H$_{c1}$ until saturation at H$_{c2}$. In many cases, the phase boundary at H$_{c1}$ satisfies a power law of the form:
T$_c\propto(H-H_{c1})^{2/d}$ (d: dimensionality), which corresponds to a Bose-Einstein condensation (BEC) of triplons universality 
class\cite{08_giamarchi}. This type of phase transition generates a staggered magnetization transverse to the external 
field, creating a canted antiferromagnetic state in the intermediate regime between H$_{c1}$ and H$_{c2}$. 
This behaviour has been observed in BaCuSi$_2$O$_6$ (Ref.~[7]), TlCuCl$_3$(Ref.~[8]), and NiCl$_2$-4SC(NH$_2$)$_2$(Ref.~[9]). An important 
property of these systems is that they must possess U(1) rotational symmetry as required by BEC theory\cite{99_giamarchi}, or at the very
least the anisotropic spin terms must be small enough that they do not alter the universality class of the phase transition.  

Recently, a new class of spin dimer compounds have been discovered with the general formula 
A$_3$M$_2$O$_8$ (Refs.~[11-13]) where A~$=$~Ba or Sr and M~$=$~Cr or Mn. At room temperature, these compounds crystallize 
in the R-3m space group, and the crystal structure consists of MO$^{4-}$ 
tetrahedra and isolated A$^{2+}$ ions. The magnetic M$^{5+}$ ions may carry spins of either S~$=$~1/2 or 1, and these are 
arranged in double-stacked triangular lattices with three-fold periodicity and so form dimers along the c-axis. These 
systems are all described well by interacting dimer models, and so they provide a new opportunity to study field-induced 
quantum phase transitions. 

In this work, we focus on the particular S~$=$~1/2 system Ba$_3$Cr$_2$O$_8$. We have completed magnetic torque, 
magnetocaloric effect(MCE), and specific heat measurements at the National High Magnetic Field Laboratory (NHMFL) on single crystals 
1-5 mg in size. We have also performed magnetic force measurements at NHMFL, using a very small $\sim1.3$~$\mu$g sample. These 
measurements allowed us to map out the phase diagram for Ba$_3$Cr$_2$O$_8$ and to investigate the associated quantum phase 
transitions. We observed only two phase transitions as a function of field, and so we found no evidence for magnetization plateaus between
H$_{c1}$ and H$_{c2}$ in our torque and force measurements. We also found that the lower phase transition can likely be described by BEC triplon 
theory, while the upper phase transition appears to be first order. 

\begin{figure}[t]
\includegraphics[width=3in,angle=0]{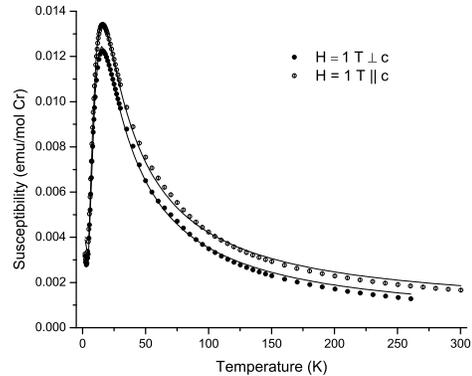}
\caption{\label{fig1}
DC susceptibility measurements of Ba$_3$Cr$_2$O$_8$ with an applied field of 1~T.}
\end{figure}

Single crystals of Ba$_3$Cr$_2$O$_8$ were grown by the travelling solvent floating zone method\cite{08_aczel}, and a detailed room temperature structure 
determination was completed as described in Ref.~[15]. 
Fig.~\ref{fig1} depicts DC susceptibility measurements of our crystals for two different orientations. The large drop with 
decreasing temperature is characteristic of systems with non-magnetic spin-singlet ground states, and the small difference between the two curves
indicates that this system has a nearly isotropic g-tensor. In accordance with previous work\cite{06_nakajima}, we fit the data to an interacting dimer 
model:
\begin{equation}
\chi_M=\frac{N_A(\mu_Bg)^2}{k_BT(3+\exp(J_0/T)+J'/T)}+\chi_0+\frac{A}{T}
\end{equation}
where N$_A$ is Avogadro's number, $\mu_B$ is the Bohr magneton, J$_0$ is the intradimer exchange constant, and J' is the sum of the interdimer 
exchange constants. The exact arrangement of the exchange constants is described elsewhere\cite{08_kofu}. The last two terms represent susceptibility 
contributions from Van Vleck paramagnetism/core diamagnetism and impurity/defect spins respectively. In principle, this fitting method can be employed 
with g, J$_0$, and J' all as separate fitting parameters. However, the fits are generally insensitive to the precise value of J' and furthermore, J' and 
g tend to trade off with one another. We note here that the exchange couplings J and J' have been determined by recent inelastic neutron scattering 
measurements to be J$_0\sim$27.6(2)~K and $|$J'$|\le$6.0(2)~K respectively\cite{08_kofu}, and recent electron spin resonance measurements (ESR)\cite{09_kofu} 
suggest the orientation-dependent g-factors are less than 2. These conditions put serious constraints on our fits, and we find that they are consistent 
with our susceptibility data only if J'$<$0 (ferromagnetic). For example, if we fix J'~$=$~-6~K in our fitting, then we find J$_0\sim$25.2(1)~K, 
g$_{ab}\sim$1.94(1), g$_c\sim$1.95(1), and a Curie constant which corresponds to only $\sim$1$\%$ free Cr$^{5+}$ spins.

Magnetic torque was measured as a function of applied field in a resistive magnet at NHMFL. The crystals were offset 
by a few degrees from the H$\parallel\hat{c}$ orientation so that it would be possible to measure a non-zero torque. A 
representative plot of torque/field ($\propto$ magnetization) vs. field at 600~mK is depicted in Fig. 2(a). For low fields, 
only a small torque is measured as we are essentially in a non-magnetic state. However, for H $>$ 12.70~T there is an abrupt
upturn in the torque due to a strong anisotropy that develops in the suspectibility tensor of the system. This behaviour is 
consistent with what one would expect for a magnetically-ordered state. Torque/field then proceeds to increase almost linearly 
up to the saturation field of $\sim$23.37~T. Note that these two critical fields were determined by finding extrema in the second 
derivative of torque/field (inset of Fig. 2(a)), similarly to what has been done in other cases\cite{05_sebastian, 08_samulon}. 
The presence of only two critical fields and the lack of magnetization plateaus in our data suggests that the triplons are highly 
delocalized in Ba$_3$Cr$_2$O$_8$ and the kinetic energy terms dominate in the relevant microscopic Hamiltonian. These observations 
are consistent with the possibility that the lower transition can be described by BEC triplon theory. 

An additional feature of the data that is particularly interesting is the magnetic hysteresis observed in association with the 
upper transition at H$_{c2}$. This suggests that while the lower transition is likely of a second-order nature, the upper 
transition is more first-order-like and lattice coupling may play a crucial role there.

Magnetic force measurements were also performed in a resistive magnet at NHMFL using a Faraday balance micromechanical 
magnetometer\cite{08_chan}. One advantage of this method over magnetic torque is that one can work with very small samples 
($\sim1$~$\mu$g). A plot of the resulting magnetization vs. field at 600~mK is depicted in Fig. 2(b), and it is 
apparent that the main qualitative features are in agreement with the torque measurement, including the magnetic hysteresis observed
near the upper transition. Note that the critical fields were determined in an analogous way to the torque measurements - by locating 
extrema in the second derivative of the magnetization. 

\begin{figure}[t]
\includegraphics[width=3in,angle=0]{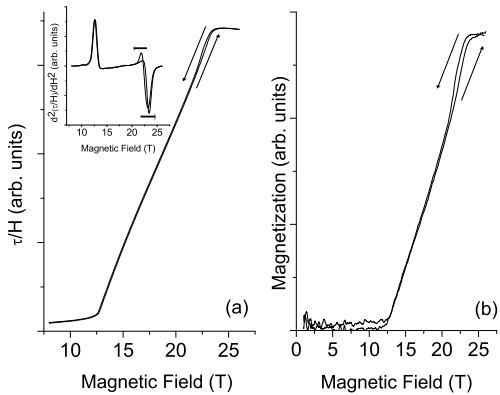}
\caption{\label{fig2}
(a) Magnetic torque measurement at 600 mK. The second derivative of torque/field, shown in the inset, 
shows two prominent extrema, indicative of the two transitions. 
(b) Force magnetometry measurement at 600 mK. This data is qualitatively similar to the torque measurement.}
\end{figure}

Further details of the phase transitions were uncovered by performing MCE and heat capacity measurements in a resistive 
magnet at NHMFL using a home-built calorimeter. All measurements were performed with H$\parallel\hat{c}$. Some 
representative MCE scans are shown in Fig. 3(a) and (b), for cases of sweeping the field both up (dotted lines) and down
(solid lines) at 2~T/min. Since the MCE is a quasi-adiabatic process, an abrupt change in the sample temperature is observed 
upon crossing an order-disorder transition to ensure entropy conservation\cite{06_silhanek}. If the phase transition is 
second order, this process should be reversible - the temperature will increase (decrease) by the same amount upon entering 
(leaving) the ordered state. However, if the phase transition is first order this will introduce a dissipative component to the 
temperature change that is always positive, and so the MCE will become irreversible. In the present case, we find the MCE 
traces are essentially reversible at the lower transition, but are highly irreversible at the upper transition, especially
for lower temperatures. This provides further evidence that the lower transition is of a second-order nature, while the 
upper transition is first order. Note that the transition points were found by locating extrema in the first derivative of 
T(H) in a similar way to what has been done previously for other systems\cite{05_sebastian, 06_zapf}.

Heat capacity measurements are shown for selected applied fields in Fig. 3(c) and (d). Both the standard thermal relaxation method
(for 13, 22, and 23~T) and the dual slope method\cite{86_riegel} (for 15, 18, and 20~T) were used to estimate the heat 
capacity. In all cases, the zero field background contribution was subtracted. A large lambda anomaly is observed in the 
intermediate field cases, but as the field is decreased closer to H$_{c1}$ or increased closer to H$_{c2}$ this becomes much
less prominent and the magnitude of the heat capacity drops off sharply. Finally, although the anomaly remains distinctly lambda-like 
even down to 13~T, at 23~T the anomaly starts to look more symmetric. This suggests that the phase transition becomes more 
first-order-like in this region, and is consistent with our other measurements.

\begin{figure}[t]
\includegraphics[width=3.5in,angle=0]{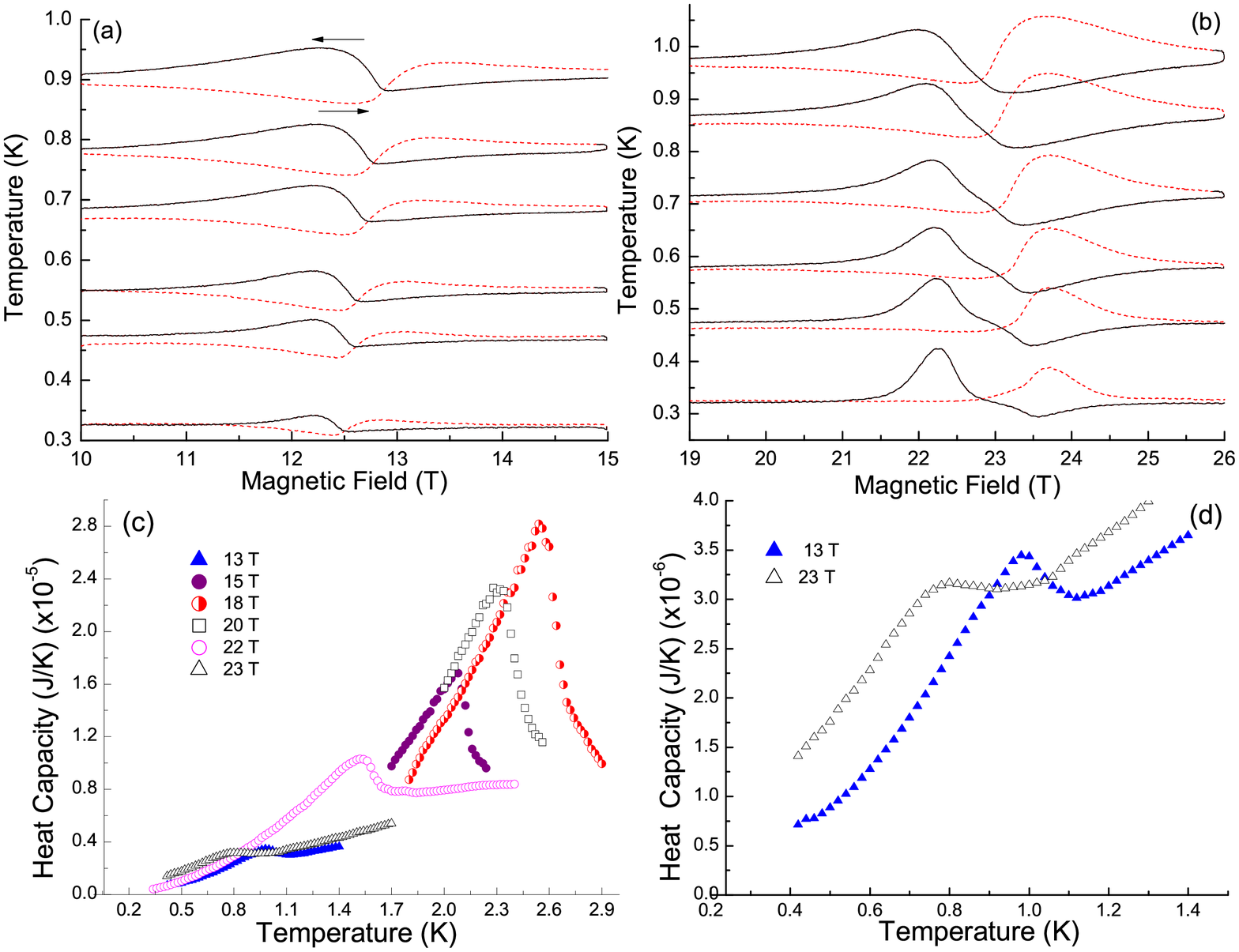}
\caption{\label{fig3}(color online)
(a),(b) MCE measurements showing the lower (upper) transition in Ba$_3$Cr$_2$O$_8$ by sweeping the field both up (dotted lines) and 
down (solid lines). 
(c) Heat capacity measurements for selected fields. 
(d) A close-up of the 13~T and 23~T heat capacity curves.}
\end{figure}

Fig. 4 combines our results in a phase diagram. We attribute the small discrepancies in the phase boundaries amongst the various
techniques to slight differences in sample orientation. The maximum transition temperature to the magnetically-ordered state was found 
to correspond to $\sim$2.7~K at H$\sim$18~T. The phase diagram is also very nearly symmetric, as is expected for a system with a much larger intradimer 
than interdimer interaction (i.e. the present case), and is due to particle-hole symmetry that comes about from the effective Hamiltonian 
describing these systems\cite{04_jaime}. The small asymmetry in the phase diagram is likely due to lattice coupling associated with the upper phase 
transition and possible contributions from the S$_z=$~0 and -1 triplet states. The latter are neglected in the aforementioned Hamiltonian. 

Dzyaloshinskii-Moriya (DM) interactions are often the source of spin anisotropy that leads to U(1) symmetry breaking and hence non-BEC behaviour in 
spin dimer systems. While recent elastic neutron scattering measurements\cite{08_kofu} have found 
evidence for a structural phase transition in Ba$_3$Cr$_2$O$_8$ at $\sim$70~K from the highly symmetric space group R-3m to the monoclinic space group 
C2/c, it turns out that the dimer centers are inversion centers in both the low and high temperature structures of this material. This suggests that 
intradimer DM interactions should be negligible in Ba$_3$Cr$_2$O$_8$. However, ESR 
measurements\cite{09_kofu} were performed very recently to study the issue of spin anisotropy in Ba$_3$Cr$_2$O$_8$ 
further, and evidence was found for a weak DM interaction of less than 0.1 meV. The detection of this weak DM interaction suggests either that it is an 
interdimer effect or that the low temperature crystal structure of Ba$_3$Cr$_2$O$_8$ has an even lower symmetry than that of the space group C2/c. In any 
event, as a result of this finding the most important issue to address is whether the resulting spin anisotropy is negligible in the sense that the lower 
transition can still be described by BEC theory.     

\begin{figure}[t]
\includegraphics[width=3.2in,angle=0]{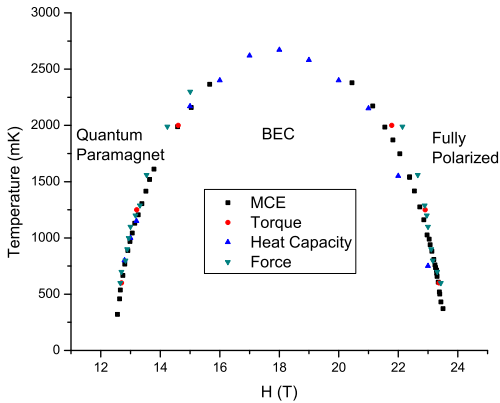}
\caption{\label{fig4}(color online)
The phase diagram of Ba$_3$Cr$_2$O$_8$ for the H$\parallel$$\hat{c}$ orientation.}
\end{figure}

The best way to verify that the spin anisotropy is negligible is to ensure that the lower phase boundary obeys a power law of the form: 
T$_c\propto(H-H_{c1})^\nu$ with $\nu~=$~2/d. With this in mind, we calculated the critical exponent pertaining to our lower transition to determine 
whether or not our data satisfied this criterion. More specifically, we used a windowing analysis technique first introduced in Ref.~[18]. Our 
narrowest fitting window contained data points in the range 333~mK~$\le$~T~$\le$~891~mK and yielded a critical exponent of 0.49(2). This is substantially 
different from any exponent pertaining to a BEC universality class, and actually agrees with the expected exponent of 0.5 corresponding to the Ising 
universality class for easy-axis magnetic systems. However, determining accurate critical exponents reliably is often quite tricky, as one needs to 
ensure that the experimental data lies in the universal regime. To this end, recent elastic neutron scattering and heat capacity measurements\cite{09_kofu}
have determined the lower phase boundary down to 30~mK and a power law fit with $\nu~=$~2/3 seems to reproduce the data well up to $\sim$1~K, suggesting 
that the lower transition is still well-described by BEC even in the presence of weak DM interactions. This is further supported by the observation of a 
canted antiferromagnetic state and gapless Goldstone mode in the neutron measurements for H $>$ H$_{c1}$. Regarding the latter, significant anisotropic 
spin contributions would instead result in an ordered state between H$_{c1}$ and H$_{c2}$ with a gapped excitation mode\cite{06_sebastian_2}.  

In summary, we have determined the phase diagram for Ba$_3$Cr$_2$O$_8$ through a combination of magnetic torque, magnetic force, MCE, and heat capacity 
measurements. We have found evidence for only two field-induced quantum phase transitions in this system, as there are no magnetization plateaus in our 
torque and force magnetometry data for H$_{c1}$$<$H$<$H$_{c2}$. The lower transition appears to be second order and well-described by BEC theory, while 
the upper transition appears to be first order. The role of lattice involvement in the upper transition and how this may be incorporated within the framework
of BEC theory remains an open question. ESR measurements performed in high magnetic fields near H$_{c2}$ may be able to shed some light on this issue.        
\\
\begin{acknowledgments}
We acknowledge useful discussions with C.D.~Batista and technical assistance from A.B. Dabkowski and E.S. Choi. We appreciate the hospitality at NHMFL 
where the majority of these experiments were performed. A.A.A. is supported by NSERC CGS, and research at McMaster University is supported 
by NSERC and CIFAR. K.N., H.B.C., and L.B. are supported by NHMFL UCGP.
\\
\end{acknowledgments}

\vfill \eject
\end{document}